\documentstyle[preprint,aps]{revtex}
\title{Data Compression of Quantum Code}

\author{K.Y.\ Szeto\\ \ \\
School of Natural Science\\
Institute for Advanced Study, \\
Olden Lane, Princeton, NJ08540, USA \\
  and \\
Dept. of Physics, \\
The Hong Kong University of Science and Technology,\\
Clear Water Bay, Kowloon, Hong Kong.\\
Email: PHSZETO@USTHK.UST.HK}

\begin{document}
\maketitle
\thispagestyle{empty}
\begin{abstract}
\end{abstract}

\noindent{1996 PACS numbers: 03.65.Bz, 05.30.-d, 89.70.+c}

\section{Introduction}

The generalization of 
the Shannon's noiseless coding theorem\cite{shannon1} 
to the case of quantum code has been performed recently 
by Schumacher\cite{schu}. 
For a quantum source $Q$ emitting states $|a_i\rangle$ 
at probability $p_i$, 
Schumacher associates with $Q$ 
the von Neumann entropy,  $S(\rho)=-Tr (\rho log \rho)$  
where $\rho =\sum_i p_i |a_i\rangle \langle a_i| $ 
is the density matrix. 
This von Neumann entropy\cite{neumann} 
plays the role of Shannon entropy. 
Soon after this work, Josza and Schumacher\cite{josza-schu} 
propose a simpler proof of the quantum noiseless theorem 
and provide a specific algorithm of data compression for quantum signals.
The general fidelity limit for quantum channels is recently 
given by Barnum et al\cite{barnum}.  
In all these works, the focus is on 
independent identically distributed quantum signals which 
comes from an irreducible Hilbert space. The transmission 
of pure quantum states $|a_i\rangle$ of the system $Q$ with 
probability $p_i$ is encoded by some state $W_i$ of the 
channel C and delivered to a receiver who decodes the signal 
and obtain a state $w_i$ of $Q$. 
The key results of these works concern the fidelity $\bar F$
of the received signal $w_i$ with respect to the signal 
source, 

\begin{equation} 
\bar F = \sum_i p_i Tr(|a_i\rangle \langle a_i| w_i) . 
\end{equation}
The quantum noiseless theorem states that 
for given $\epsilon,\delta >0,$ and 
a given channel with $S(\rho)+\delta$ 
qubits available  per input state, then 
for all sufficiently large N, there exists a coding and a 
decoding scheme which transmits blocks of N states with 
average fidelity $\bar F > 1-\epsilon$. 
They also prove the converse of the theorem which states that 
for given $\epsilon,\delta >0,$ and 
a given channel with  $S(\rho)-\delta$ qubits available 
per input state, then for all sufficiently large N, 
for any coding and decoding scheme for blocks of N states, 
the average fidelity satisfies $\bar F <  \epsilon$. 

The Josza-Schumacher scheme of data compression 
provides a convenient way of forming block codes with N states 
for a given source $Q$ which Hilbert space has a 
dimension $d$. Here I make two simple observations:  
(1) what happens if the Hilbert space of $Q$ can be decomposed 
into two or more mutually orthogonal subspaces, and that the 
signals state $|a_i\rangle$ belongs to only one of these 
subspaces? Can we do something simpler and easier than 
the Josza-Schumacher scheme? 
(2) Given an irreducible Hilbert space of dimension $d$, 
a quantum coding device that use $q$-ary quantum code unit, 
(analogous to the $q$-ary alphabet in classical coding), 
and certain limit on the size $N$  of the block code, 
is there a simple relation on $(d,q,N)$ that allows  data compression 
with the least amount of wasteful resource? 
I show that for (1) there is an alternative way of 
coding and decoding that achieves the same quality of 
data compression, but employs classical data compression 
together with  block quantum code. This will allow a more 
familiar method of tackling the quantum signals and use 
less resource, assuming that classical data compression is 
easier and cheaper than its quantum counterpart. 
As for (2), I have derived a simple rule of thumb for resource 
allocation in the Josza-Schumacher scheme. 
Some numerical solutions which do not waste any resource  
are tabulated. 
 
\section{Block Code for Decomposable Hilbert Space}

Imagine a quantum source which emits signals 
according to certain known 
selection rules. Hence one can  
separate the total Hilbert space of the 
signals into two or more orthogonal subspaces. An example 
is the Bell basis where the quantum signals are generated 
by linear combination of the following states:
$(\Psi^-, \Psi^+, \phi^+, \phi^-)$ , 

\begin{equation}
\Psi^{\pm} = {1\over {\sqrt 2}} 
\biggl(|\uparrow \downarrow\rangle \pm 
|\downarrow \uparrow\rangle \biggr), 
\end{equation}

and 

\begin{equation}
\phi^{\pm} = {1\over {\sqrt 2}} 
\biggl(|\uparrow \uparrow\rangle \pm 
|\downarrow \downarrow\rangle \biggr). 
\end{equation}
In this case, the state $\Psi^-$ is the singlet and spans a 
Hilbert space $H_1$ of dimension $d_1 =1$, and the states 
$(\Psi^+, \phi^+, \phi^-)$ form the triplet and span a 
Hilbert space $H_2$ of dimension $d_2 =3$. 
This is an example where spin $1/2$ signal states can be 
sent from the source $Q$ in a coherent manner so that 
the signals are of two possible kinds, pair of signals 
can be taken from either $H_1$ or $H_2$, but not mixed. 
More generally, if the signal states can be written either 
in the form 

\begin{equation}
|a\rangle = \sum_i \alpha_i |e_1^i\rangle 
\end{equation} 

or 

\begin{equation}
|b\rangle = \sum_i \beta_i |e_2^i\rangle , 
\end{equation} 
with $\{ |e_k^i\rangle , i=1,..,d_k\}$ an orthonormal 
basis for $H_k$ for $k=1$ or $2$ and $H_1$ is orthogonal to $H_2$, 
then we can write the density matrix for the source $Q$ as

\begin{equation}
\rho = P_1 \rho_1 + P_2 \rho_2
\end{equation}
where $P_1$ is the probability that the signal comes from $H_1$ 
and $P_2=1-P_1$ is the probability that it comes from $H_2$.  
We can represent $\rho_1=\sum_j q_2^j |a_j\rangle\langle a_j|$ 
and $\rho_2=\sum_j q_1^j |b_j\rangle\langle b_j|$, where 
$\{ q_k^j\}$ are the probabilities of the signal states given 
that they are from the Hilbert space $H_k$. We also introduce the 
projection operator $\Pi_1 =\sum_j |e_1^j\rangle \langle e_1^j|$ 
that will be useful later to construct the N-block classical code. 
For a sequence of N signal states  
$\{|c_1\rangle,..,|c_N\rangle\}$, the operation of $\Pi_1$ yields 
a binary string $\vec x=(x_1,x_2,..,x_N)$ of length N, with 
each ${x_i = 0}$ if ${|c_i\rangle \ \epsilon \ H_1}$ and  
${x_i = 1}$ if ${|c_i\rangle \ \epsilon \ H_2}$.  
We now show that the von Neumann entropy 
for the full density matrix $\rho$ can be decomposed 
into three terms,

\begin{equation}
S(\rho )= 
H(X) + P_1 S(\rho_1 )+ P_2 S(\rho_2 )
\label{sentropy}
\end{equation} 
where $H(X)=-P_1 log(P_1) -P_2 log(P_2)$ is the Shannon entropy 
associated with the classical code $X=\{ \vec x \}$. 
(This result is the quantum version of the classical result 
$H(X,Y)=H(X)+H(Y|X)$\cite{ash}.) 
Eq.\ref{sentropy} can be easily obtained by noting that 

\begin{equation}
Tr(P_1 \rho_1 + P_2 \rho_2)log (P_1 \rho_1 + P_2 \rho_2)) 
= Tr (P_1 \rho_1 log(P_1 \rho_1) + P_2 \rho_2 log(P_2 \rho_2))
\end{equation}
as $Tr(\rho_1 log(\rho_2))= Tr(\rho_2 log(\rho_1))=0$,  
since $H_1$ is orthogonal complement of $H_2$. 
Eq.\ref{sentropy} relates two methods for 
constructing block codes using the quantum noiseless coding theorem. 
The first method is the direct  
application of the Josza-Schumacher technique to 
the entire Hilbert space $\cal H$ without taking advantage 
of the decomposition of $\cal H$ into two orthogonal subspaces, 
and this will take up a channel resource of at least 
$S(\rho)$ qubits. 
A second method is to make use of the right hand side of 
eq.\ref{sentropy}. 
We  first perform a classical block code using for example 
the Huffmann coding technique\cite{huffmann}
on the $N$-sequence $\{ \vec x \}$, and this will 
take up at least 
$H(X)$ bit per input of classical channel resource, 
then one can perform respectively the quantum block coding 
for the $N_1$-subsequence of the states from $H_1$ 
and use up at least 
$S(\rho_1)$ qubits of quantum channel resource per 
$H_1$ input state, 
and similarly for the $N_2$-subsequence of the states from $H_2$. 
The lower limit of 
total channel resource is the same as given by eq.\ref{sentropy}, 
but the advantage of the second method is that by making the most use 
of the information of the Hilbert space structure, one can 
perform classical block coding of $X$ before 
one performs the quantum block coding of 
$|Y\rangle=|a_1\rangle|a_2\rangle...|a_{N_1}\rangle$ 
and of 
$|Z\rangle=|b_1\rangle|b_2\rangle...|b_{N_2}\rangle$ . 
This replaces part of the quantum resource by 
classical resource $H(X)$. 
Since classical coding is easier than quantum coding, 
one should first inquire if the Hilbert space of the signal states 
can be decomposed into several orthogonal subspaces before 
proceeding directly to use quantum code. 
One should note that it is necessary that the two subspaces 
$H_1$ and $H_2$ are orthogonal. In fact, one can show that block 
coding cannot separate two nonorthogonal subspaces.

\section{Parameterization of Quantum Block Code in the 
Josza-Schumacher Scheme}

The Josza-Schumacher quantum block code provides a simple 
scheme of maximizing the fidelity while tolerating small errors 
in the signals reconstituted from the coded version. 
Of course, in any realistic calculation of fidelity, 
one has to take into account the details of 
the probabilities of the states from the signal source, therefore 
a general statement on the efficient use of block code of size $N$, 
given that the dimension of the Hilbert space is $d$, 
seems impractical. 
However, we can pose the problem of 
minimizing the resource allocation of quantum code for a 
{\it particular} scheme of coding, which 
parametrization  of channel resource 
is of some utility. The scheme we discuss is a generalization of 
the Josza-Schumacher scheme, and the key idea lies in the 
observation that the quantum bits required to code the 
states in the typical $N$-sequence space of dimension 
$D_\Lambda$ usually requires more resource than necessary,  
in that a block of $q$-ary code of length $M$ in general has 
$q^M > D_\Lambda$. 

Consider block code of length $N$. The idea is to 
construct a typical subspace ${\cal L}^N$ of $N$-sequence 
of the Hilbert space ${\cal H}^N$ 
of the $N$-sequence of  signal states 
so that for any subspace with the same dimension as 
${\cal L}^N$, the fidelity will be smaller.  
Let's consider the signal states $\{|a_i\rangle, i=1,..,d^\star\}$ 
occurring with probability $\{p_i \} $.  
Without loss of generality,  we assume that the signal states 
are linearly independent but not necessary 
orthogonal, so that the space $\cal H$ spanned by them 
also has dimension $d^\star$ and 
we order $\{p_i\}$ so that $p_i \ge p_j$ for $i < j$. 
For simplicity, let's assume that there is a state $|a_d\rangle$ 
which probability $p_d$ is different from all the other states.  
(This is not necessary for the argument that follows.) 
Let ${{\cal L}=span\{ |a_1\rangle, |a_2\rangle,..,|a_d\rangle\}}$ 
be the subspace of $\cal H$. 
After defining $\cal L$, we can choose an orthonormal basis 
of $\cal L$ to be $\{ |e_1\rangle,..,|e_d\rangle\}$ with 
$|e_1\rangle \equiv |a_1\rangle$. 
Also, we extend this basis by adding an orthonormal set 
consisting of $(d^\star-d)$  vectors 
$\{ |e_{d+1}\rangle,..,|e_{d^\star}\rangle\}$ to form the basis 
of $\cal H$. 
Now consider product state 
$|\lambda_1\lambda_2..\lambda_N\rangle$ 
for a given $N$, and $|\lambda_i\rangle$ is chosen from 
the basis $\{ |e_1\rangle,..,|e_d\rangle\}$. 
We first observe that a similar product state 
$|\mu_1\mu_2..\mu_N\rangle$ with $|\mu_i\rangle$ chosen from the 
basis $\{ |e_1\rangle,..,|e_{d^\star}\rangle\}$ will generally 
has a smaller contribution to the fidelity calculation. 
(Indeed, one can verify easily that for 
given $|\mu_1\mu_2..\mu_N\rangle$, we can replace those entries 
which do not belong to  $\cal L$ by some elements in $\cal L$ 
and the new product state will give a higher 
contribution to the fidelity.)  
Let's now compare the typical subspace $\Lambda$ of ${\cal H}^N$ 
formed by a set of $D_\Lambda$ of states of form 
$|\lambda_1\lambda_2..\lambda_N\rangle$, and a similar subspace 
$\Gamma$ of ${\cal H}^N$ 
formed by a set of $D_\Gamma$ of states of form 
$|\mu_1\mu_2..\mu_N\rangle$. If we insist that these two  
subspaces of typical $N$-sequence has the same dimension, 
$D_\Lambda=D_\Gamma$, then we can show that the fidelity
${F_\lambda=\sum_{\{|a...a\rangle\}}Tr(\pi W_\lambda)}$ 
calculated 
using the projection 
${
W_\lambda \equiv \sum_{\{ \lambda \} } 
|\lambda_1\lambda_2..\lambda_N\rangle
\langle\lambda_N...\lambda_2\lambda_1|}$
will be higher than the fidelity 
${F_\mu=\sum_{\{|a...a\rangle\}}Tr(\pi W_\mu)}$ using 
${
W_\mu \equiv \sum_{\{ \mu \} } 
|\mu_1\mu_2..\mu_N\rangle
\langle\mu_N...\mu_2\mu_1|}$. 
Here 
${
\pi \equiv \sum_{a} P(a) 
|a_{i_1}...a_{i_N}\rangle\langle a_{i_N}...a_{i_1}|
}$ 
with the sum over all possible $N$-sequence $i_1....i_N$. 
Thus, the $N$-sequence chosen from $\lambda$ 
gives a control on the fidelity which is now determined 
by the integer parameters $d$ and $N$. 
One can now discuss some general results on the 
parametrization of the quantum block coding scheme using the 
product space 
${
\Lambda\equiv {\cal L}^N =
\overbrace{{\cal L} \bigotimes {\cal L} .. \bigotimes {\cal L}}^N
}$. 
It is a  typical subspace of the Hilbert space 
${
{\cal H}^N 
=\overbrace{{\cal H}\bigotimes {\cal H} .. \bigotimes {\cal H}}^N
}$   
and it contains the likely $N$-sequence of signal state. 
In order to encode these likely $N$-sequence of signal state, 
we have to calculate the dimension $D_\Lambda$ of $\Lambda$ and 
use a $q$-ary quantum code to represent the states in $\Lambda$. 

To simplify notation, we will use $|s\rangle$ to denote 
$|e_1\rangle =|a_1\rangle$ and generically $|r\rangle$ to denote 
any other state in $\cal L$. 
The particular scheme 
of quantum coding is 
to form block code of $N$ states composed of product of $K$ 
$|s\rangle$ states and $(N-K)$ $|r\rangle$ states which are 
occuring less frequently. 
The set of such product states can be written in the general form 

\begin{eqnarray*}
|ssss....s\rangle, \cr
|rss.....s\rangle, & |srss....s\rangle, &...,\ |ssss...sr\rangle, \cr
|rrs.....s\rangle, & |rsrs....s\rangle, &...,\ |ssss...rr\rangle, \cr
....., 
\end{eqnarray*}
These $N$ product states have the unique feature that one can 
unambiguously conclude that the $|s\rangle$ state is the 
majority species in the product. If there are $K (\ge N/2)$ states 
are $|s\rangle$, then the remaining states in the product 
can be selected from any signal states $|r\rangle$ from $\cal L$.  
In order to enumerate all the states and count the dimension of 
$\Lambda$, we observe that 
there is only one $|ssss...s\rangle$ state consists of product of 
$N$ $|s\rangle$ state. 
There are $(d-1)*N$ states of the 
form $|sss..srs...s\rangle$ since there are $(d-1)$ choices of 
different signal states to put into $N$ different spots in the 
string. 
In general, if there are $K$ $|s\rangle$ in the 
product $|\lambda_1\lambda_2.....\lambda_N\rangle$, 
the remaining $(N-K)$ signal states are chosen from $(d-1)$ choices 
of $|r\rangle$ states. The possible combination is 
$(d-1)^{N-K} C^N_{N-K}$ with 
$C^N_{N-K} \equiv {{N!}\over {K!(N-K)!}}$ being the number of 
ways of selecting  $K$ positions for the $|r\rangle$ states 
out of $N$ slots. If $N(=2L)$ is even and there are $L$ $|s\rangle$ 
states in the product state already, then the remaining 
$L$ slot cannot be all of the same state $|r\rangle$ in order 
to prevent ambiguity of the majority signal state, 
we then in this special case have only 
$(d-1)^{L-1}(d-2) C^N_L$ possible combination, as there 
are $(d-1)$ choices to choose the first $|r\rangle$ state, and 
$(d-2)$ choices for the second $|r\rangle$, 
since the second $|r\rangle$ state must be different from the 
first so that one knows that the majority in the product is 
$|s\rangle$. 
Summarizing this discussion, we arrive at the dimension 
$D_\Lambda$of the subspace $\Lambda$, for odd $N=2L-1$

\begin{equation}
 D_\Lambda=1 + (d-1)C^N_{N-1} + (d-1)^2 C^N_{N-2} +.. +
(d-1)^{L-1} C^N_{N-L+1} 
\end{equation} 
and for even $N=2L$, 

\begin{equation}
 D_\Lambda=1 + (d-1)C^N_{N-1} + (d-1)^2 C^N_{N-2} +.. +
(d-1)^{L-1} C^N_{N-L+1} +(d-1)^{L-1} (d-2) C^N_L. 
\end{equation} 
We note that in general 
$D_\Lambda \ll dim({\cal H}^N) = {d^\star}^N$ and the states in 
$\Lambda$ has a relatively high probability of occurence 
among the states in ${\cal H}^N$. The exact calculation of the 
probability of occurence of the states in $\Lambda$ 
requires a knowledge of probability of 
the occurence of the single signal state from 
the  quantum source. 
However, based on this general scheme of construction of 
$\Lambda$, we can say something about the optimal method of 
coding the states in $\Lambda$ with quantum code. 

Generally, quantum code makes use of qubits, equivalent 
to the minimal quantum space spanned by spin $1/2$, 
which corresponds  
to the classical binary system with the number $q$ of alphabets 
being 2. We can in general consider a spin $J$ quantum system 
which has a Hilbert space of dimension $q=2J+1$, corresponding 
to the classical $q$-ary system with $q$ alphabets. Assuming that 
a general spin $J$ system is available for quantum coding of the 
$N$ block signal, then the space $\cal Q$ of quantum codes cosists 
of states $|q_1q_2..q_M\rangle$, and the dimension of $\cal Q$ is  
$q^M$. 
In order to ensure that the quantum codes can encode all the 
informations in $\Lambda$, it is necessary that 
$q^M \ge D_\Lambda$. The resource wasted in coding the $N$ block 
signal is measured by the percentage 
$E\equiv  {{q^M - D_\Lambda}\over {D_\Lambda}}$.   
Since $D_\Lambda$ is a function of only $d$ and $N$, we can 
look for integer solution $(d,N,q,M)$ 
for the equation $q^M = D_\Lambda$ for a physical range of 
values of $d,N,q,M$. 
We have found 9 solutions for the range 
$2\le d,q,M \le 32$ and $ 3\le N \le 32$.  
They are listed in Table.1  with 
dimension $D_\Lambda = q^M$. 

\begin{tabular}{||c|c|c|c|c||}
\hline
\multicolumn{5}{||c||}{Table.1 Exact solution of $D_\Lambda = q^M$}\\ \hline
 d &  N &  q &  M & $D_\Lambda$ \\ \hline
 2 &  3 &  2 &  2 &               4\\
 2 &  5 &  2 &  4 &              16\\
 2 &  9 &  2 &  8 &             256\\
 2 & 17 &  2 & 16 &           65536\\
 2 &  5 &  4 &  2 &              16\\
 2 &  9 &  4 &  4 &             256\\
 2 & 17 &  4 &  8 &           65536\\
 2 & 11 & 32 &  2 &            1024\\
 2 & 21 & 32 &  4 &              1048576\\ \hline
 4 &  4 &  7 &  2 &                   49\\ \hline
 6 &  3 &  2 &  4 &                   16\\  
 6 &  3 &  4 &  2 &                   16\\ \hline
17 &  3 &  7 &  2 &                   49\\ \hline
22 &  3 &  2 &  6 &                   64\\ 
22 &  3 &  4 &  3 &                   64\\ 
22 &  3 &  8 &  2 &                   64\\ \hline
\end{tabular}

\section{Discussion}

We have demonstrated that the data compression for quantum signals 
can be simplified in the case where the Hilbert space of the 
signals states can be decomposed into two or more mutually orthogonal 
subspaces, as one can first perform data compression 
on the classical code encoding the particular subspace to which the 
signal state belongs before performing quantum block coding. 
The problem of performing the projection into particular subspace 
without destroying the signals depend on the quantum source, but 
in principle this can be done. (For the separation of 
horizontally polarized photons from vertically polarized ones, 
a calcite crystal can be used as the projection operator\cite{bbe}.)  
The classical signals $X$ carrying the information for 
the subspace can be block coded to achieve optimal data compression, 
and the set of quantum signals which forms block of length $N_1=P_1 N$ 
and $N_2=N-N_1$ can be coded, using for example the Josza-Schumacher 
scheme. 
Since a particular sequence of $N$ quantum signals need not 
have exactly $P_1 N$ signals from $H_1$, one should use some 
extra dimension to encode the signals from $H_1$. This will be 
a disadvantage compared to the straight forward coding of the 
$N$ signals using the entire Hilbert space $\cal H$.  
In general, 
the ease of classical data compression, such as the use of 
Huffmann code, outweights the small extra dimensions needed to 
encode the string of $N_k$ quantum signals from $H_k$. 
One can consider instantaneous code, or more generally 
uniquely decipherable code for quantum signals in the context 
of the present discussion. The fact that $P_1 N$ signals 
from $H_1$ in general has some fluctuation suggests that one 
can form hierarchical block of the quantum code 
$|h_1h_2...h_{M_1}\rangle$ with ${h_i \epsilon\{0,1,..,q_1-1\}}$ 
where $q_1^{M_1} \ge D_{\Lambda_1}$ with $\Lambda_1$ being the 
typical subspace of $N_1$ sequence of the Hilbert space 
${H_1}^N$.

Finally, the possibility of decomposing signals belonging to 
$\cal H$ into signals belonging to different orthogonal subspaces 
$H_k$ depends on the quantum source. As the example of the 
Bell basis demonstrates, one may anticipate quantum source 
emitting signals with certain selection rule, 
or with certain quantum correlation which renders the 
Hilbert space $\cal H$ decomposable. This naturally leads to 
the question of quantum source emitting 
correlated quantum signals and poses an interesting problem 
for future research. For now,  if the signals are mostly 
from either one or the other orthogonal subspace, and only a 
tiny fraction of the signals form linear combination of states 
from two different subspaces, one can still employ the technique 
discussed in this paper in focusing on the typical sequence, 
after discarding the signals that are mixtures of two subspaces, 
but at the price of a compromised fidelity.

\section{Ackowledgement}

I acknowledge many helpful discussion with Hoi-Kwong Lo and 
the hospitality of the School of Natural Sciences at 
the Institute for Advanced Study. Part of this 
research is funded by the Hong Kong Telecom Institute 
of Information Technology.


\begin{thebibliography}{99}

\bibitem{shannon1}
C.E. Shannon, Bell System Technical Journal {\bf 27}, 379(1948). 

\bibitem{schu}
B. Schumacher, Phys. Rev. A{\bf 51}, 2738-2747(1995). 

\bibitem{neumann}
J. von Neumann, {\it Mathematical Foundations of Quantum Mechanics} 
(English translation by R.T. Beyer), Princeton University Press (1955). 
\bibitem{josza-schu}
R.Jozsa and B. Schumacher, J. Mod. Optics{\bf 41},2343-2349(1994).

\bibitem{barnum}
H.Barnum, C.A. Fuchs, R. Jozsa, and B. Schumacher, 
preprint, quant-ph/9603014, 8 Mar 96

\bibitem{ash}
R.B. Ash; {\it Information Theory}, Dover, (1965).

\bibitem{huffmann}
D.A. Huffmann, Proc.IRE, {\bf 40}, No.10, 1098-1101(1952). 

\bibitem{bbe}
C.H. Bennett, G. Brassard, and A.K. Ekert; 
Scientific American, 50, Oct.(1992)

\end{thebibliography}
\end{document}